\documentclass[preprint,12pt]{elsarticle}

\usepackage{graphicx}

 \usepackage{epsfig}

\usepackage{amssymb}

\usepackage{lineno}
\usepackage{multirow}
\usepackage{booktabs}
\usepackage{amsbsy}

\journal{ApJ}

\begin{document}

\begin{frontmatter}


\title{Neutrino analysis of the September 2010 Crab Nebula flare and time-integrated constraints on neutrino emission from the Crab using IceCube.}
\author[Madison]{R.~Abbasi} 
\author[Gent]{Y.~Abdou} 
\author[RiverFalls]{T.~Abu-Zayyad} 
\author[Christchurch]{J.~Adams} 
\author[Madison]{J.~A.~Aguilar} 
\author[Oxford]{M.~Ahlers} 
\author[Aachen]{D.~Altmann} 
\author[Madison]{K.~Andeen} 
\author[Wuppertal]{J.~Auffenberg} 
\author[Bartol]{X.~Bai} 
% ...
\address[Aachen]{III. Physikalisches Institut, RWTH Aachen University, D-52056 Aachen, Germany}
% ...
\address[Christchurch]{Dept.~of Physics and Astronomy, University of Canterbury, Private Bag 4800, Christchurch, New Zealand}
% ...
\address[Gent]{Dept.~of Physics and Astronomy, University of Gent, B-9000 Gent, Belgium}
% ...
\address[Madison]{Dept.~of Physics, University of Wisconsin, Madison, WI 53706, USA}
% ...
\address[Bartol]{Bartol Research Institute and Department of Physics and Astronomy, University of Delaware, Newark, DE 19716, USA}
\address[Oxford]{Dept.~of Physics, University of Oxford, 1 Keble Road, Oxford OX1 3NP, UK}
\address[RiverFalls]{Dept.~of Physics, University of Wisconsin, River Falls, WI 54022, USA}
% ...
\address[Wuppertal]{Dept.~of Physics, University of Wuppertal, D-42119 Wuppertal, Germany}
% ...
\fntext[Bari]{also Universit\`a di Bari and Sezione INFN, Dipartimento di Fisica, I-70126, Bari, Italy}
\fntext[Goddard]{NASA Goddard Space Flight Center, Greenbelt, MD 20771, USA}


\begin{abstract}

  We present the results for a search of high-energy muon neutrinos with the
  IceCube detector in coincidence with the Crab nebula flare reported on
  September 2010 by various experiments. Due to the unusual flaring state of
  the otherwise steady source we performed a prompt analysis of the 79-string
  configuration data to search for neutrinos that might be emitted along with
  the observed $\gamma$-rays. We performed two different and complementary
  data selections of neutrino events in the time window of 10 days around the
  flare.  One event selection is optimized for discovery of $E_{\nu}^{-2}$
  neutrino spectrum typical of $1^{st}$ order Fermi acceleration. A similar
  event selection has also been applied to the 40-string data to derive the
  time-integrated limits to the neutrino emission from the Crab \cite{jon}.
  The other event selection was optimized for discovery of neutrino spectra
  with softer spectral index and TeV energy cut-offs as observed for various
  galactic sources in $\gamma$-rays.
  The 90\% CL best upper limits on the Crab flux during the 10 day flare are
  {$4.73 \times 10^{-11}$ cm$^{-2}$ s$^{-1}$TeV$^{-1}$} for an $E_{\nu}^{-2}$
  neutrino spectrum and {$2.50 \times 10^{-10}$ cm$^{-2}$ s$^{-1}$TeV$^{-1}$}
  for a softer neutrino spectra of $E_{\nu}^{-2.7}$, as indicated by Fermi
  measurements during the flare. IceCube has also set a time-integrated limit
  on the neutrino emission of the Crab using 375.5 days of livetime of the
  40-string configuration data. This limit is compared to existing models of
  neutrino production from the Crab and its impact on astrophysical
  parameters is discussed. The most optimistic predictions of some models are
  already rejected by the IceCube neutrino telescope with more than 90\% CL.

\end{abstract}


\end{frontmatter}


\section{Introduction}
\label{sec1}

The Crab supernova remnant, originating from a stellar explosion at a
distance of 2 kpc recorded in 1054 AD, consists of a central pulsar, a
synchrotron nebula, and a surrounding cloud of expanding thermal
ejecta~\cite{Hester2008}. Its bright and steady emission has made it a
standard candle for telescope calibration. However, the photon emission
stability in the X-ray and in the $\gamma$-ray regions is recently being
questioned by a number of satellite experiments. As a matter of fact, a 7\%
decline of the Crab flux in the 3-100 keV region, larger at higher energies,
has been observed in the period between 2008 and 2010 by the Fermi Gamma-ray
Burst monitor and confirmed by Swift/BAT, RXTE/PCA, and INTEGRAL
(IBIS)~\cite{crabby}. The pulsed emission from RXTE/PCA observations is
consistent with the observed pulsar spin-down suggesting that the decline is
due to changes in the nebula and not in the pulsar.

The source of energy that powers the Crab is the spin-down luminosity of the
pulsar. 
The measured spin-down luminosity of the pulsar is $\sim 5 \times 10^{38}$ erg
s$^{-1}$ and its rotational period is 33 ms.  While a small fraction of this
energy goes into the pulsed emission, most of it is carried by a highly
magnetized wind of relativistic plasma, the composition of which is not known. Both
pure $e^{\pm}$ plasma models and a mixture of $e^{\pm}$ and protons or ions
have been proposed ~\cite{Hester2008,Blasi,Bednarek2003,Protheroe,nureview}.
The wind terminates in a standing shock and transfers some of the energy to
accelerating particles. A part of this energy is converted into synchrotron
emission from radio to MeV $\gamma$-rays by a population of high energy
electrons radiating in the nebular magnetic field. The observations of the
synchrotron emission from the Crab up to the MeV energies, make the Crab an
undisputed galactic accelerator able to inject electrons up to energies $\sim
10^{15}$ eV.  These high energy electrons inevitably interact with the
ambient photon fields through inverse Compton scattering, resulting in the production of
high-energy $\gamma$-rays observable in the TeV
regime~\cite{HESS2006,HEGRA,Magic_crab}.  The synchrotron emission
from the Crab has an integrated luminosity of $\sim 1.3 \times 10^{38}$~erg s$^{-1}$, that is, at least $\sim$26\% of
the spin-down luminosity of the pulsar is involved in the acceleration of
electrons in the energy range $10^{11}$ -- $10^{15}$~eV~\cite{Hester2008}.  On
the other hand, the presence of hadrons in the pulsar wind and the amount of
energy transported by them remain as some of the unresolved and interesting
questions about the Crab Nebula and plerions in general.

Protons and ions do not lose their energy as efficiently as electrons, and hence it is more difficult to observe the products of their
interactions. The dominant processes, discussed below, are proton-proton and proton-$\gamma$ interactions, and
both processes generate $\gamma$-rays and neutrinos through meson decays. Hence, neutrinos
constitute an unique signature for hadron acceleration while hadronic $\gamma$-ray production has to be disentangled from inverse Compton emission.
Hadronic models of the Crab emission assume that the pulsar wind is composed of a mixture of electrons and ions. These models predict that a significant
part of the rotational energy lost by the pulsar is transferred through the
shock radius to relativistic nuclei in the pulsar wind.  Relativistic nuclei
injected into the nebula can interact with the nebula matter, and produce
cosmic rays and neutrinos via pion decay.  Neutrino production by protons and
nuclei interacting in the pulsar wind in the Crab have been discussed in
Ref.~\cite{Blasi,Bednarek2003}. According to these models, the nuclei can
generate Alfv\'en waves just above the pulsar wind shock. These Alfv\'en
waves will resonantly scatter off and accelerate the positrons and electrons that
create the synchrotron emission.  In the model
described in Ref.~\cite{Protheroe} neutrinos are produced by heavy nuclei
accelerated by the rotating neutron star that photo-disintegrate in
collisions with soft photons. These models predict between $1-5$ events per
year in a cubic-kilometer detector such as IceCube when accounting for neutrino
oscillations. Inelastic nuclear collisions are considered in
Ref.~\cite{Blasi}. In this paper the predicted rates depend on the Lorentz factor,
$\Gamma$, of nuclei injected by the pulsar and the effective target
density.  
The thermal matter distribution in the Crab is far from being uniform but forms filaments. 
For relativistic protons the effective target density is also affected by the structure of the magnetic field in and around these filaments.
The authors in Ref.~\cite{Blasi} provide several expected neutrino fluxes
from the Crab Nebula as a function of energy, for different assumptions on these two parameters.
For the highest values of the effective target density, IceCube begins to have the sensitivity to probe the
highest possible values around $\Gamma \lesssim 10^7$ while the favored
values of the upstream Lorentz factor of the wind are $\Gamma \sim
10^6$~\cite{arons}.

Acceleration of positive ions near the surface of a young rotating neutron
star ($\lesssim 10^5$ yrs) has also been investigated in Ref.~\cite{Burgio}.
This model describes how positive ions can be accelerated to $\sim 1$~PeV in
rapidly-rotating pulsars, with typical magnetic fields ($B \sim 10^{12}$~G), by
a potential drop {\it across} the magnetic field lines of the pulsar. Assuming
that the star's magnetic moment $\mu$ and the angular velocity $\Omega$
satisfy the relation ${\vec{\mu} \cdot \vec{\Omega}} < 0$, protons are
accelerated away from the stellar surface. Beamed neutrinos (in coincidence with the radio beam)
 are produced by such
high energy protons interacting with the star's radiation field when the
$\Delta$ production threshold is surpassed. Observation of these neutrinos
could validate the existence of a hadronic component and a strong magnetic
field near the stellar surface that accelerates the charged particles. The predictions in Ref.~\cite{Burgio2} based on this model account for $\sim 45$
neutrino events/yr from the Crab in a cubic-kilometer detector in the most optimistic scenario where the fraction of charge depletion is assumed to be $f_{d} \sim 1/2$. In
this paper we will show that IceCube data severely constrains these optimistic predictions of the model.

In Ref.~\cite{kappes} a mean prediction of 1.2 neutrino events per year for
$E_{\nu} > 1$~TeV was calculated for an underwater cubic-kilometer
detector. This prediction is based on the H.E.S.S. measured $\gamma$-ray
spectrum~\cite{HESS2006} assuming that all the $\gamma$-rays observed by
H.E.S.S. up to 40~TeV are produced by pion decay and that the absorption of
$\gamma$-rays is negligible. A similar calculation connecting photon and
neutrino fluxes was done in Ref.~\cite{halzen} predicting about 5 events from
the Crab accounting for neutrino oscillations. For a summary of some of the
models on neutrino spectra the reader is referred to~\cite{nureview}.

From Sep. 19 to 22, 2010 the AGILE satellite~\cite{agile_atel, agile_science} reported an
enhanced $\gamma$-ray emission above 100 MeV from the Crab nebula. The flare,
however, was not detected in X-rays by INTEGRAL~\cite{integral}
observations between Sep. 12 and 19 partially overlapping with AGILE
observations. It was also not confirmed by the SWIFT/BAT~\cite{swift} in the
15-150~keV range nor by RXTE~\cite{rxte} on a dedicated observation of the
Crab on Sep. 24.  The observation was later confirmed by the Large Area
Telescope on board of the Fermi Gamma-Ray Space Telescope that detected a
flare of $\gamma$-rays ($E_{\gamma}> 100$~MeV) with a duration of $\sim 4$
days between Sep. 19--22 in the Crab direction \cite{Fermi_flare}. The
observed energy spectrum during the flare interval was consistent with a
negative power-law with a spectral index of $-2.7 \pm 0.2$.  The flux
increase was a factor $5.5 \pm 0.8$ above the average flux from the Crab.
Fermi also detected another flare of 16 days in Feb. 2009 corresponding to a
flux increase of a factor $3.8 \pm 0.5$ but much softer spectral index ($-4.3
\pm 0.3$).  The ARGO-YBJ collaboration also issued an ATel on Sep. 2010 on
the observation of an enhancement of the TeV emission for the same period of
time but with a wider interval of 10 days. The enhanced TeV emission
corresponded to a flux about 3-4 times higher than the usual Crab flux in TeV
energies~\cite{ARGO}. However, this observation was not confirmed by
MAGIC~\cite{magic} nor VERITAS~\cite{veritas}; Imaging Cherenkov Telescopes in
a similar energy range as ARGO-YBJ. The spectral and timing properties of
the flares indicate that the $\gamma$-rays are emitted via synchrotron
radiation from PeV electrons from a region smaller than $1.4 \times
10^{-2}$~pc. This dimension is comparable to the jet knots observed close to
the termination shock of the Crab Nebula~\cite{Chandra}. Even though the
Crab has always been considered to be a source of synchrotron emission, the
flare represents a challenge to shock diffusive acceleration
theory~\cite{Fermi_flare}. Nonetheless, explanations of the high variability
due to electromagnetic phenomena have been proposed in Ref.~\cite{Bednarek}
where the emission comes from a part of the pulsar wind
shock\footnote{During the final stage of the editing of this paper another large flare was observed from the Crab~\cite{Fermi_April}. This flare is even more intense than the one observed in September and is being studied by various experiments. Hence, IceCube analysis will happen when results from Fermi, other X-ray satellites and other TeV ground based experiments will be available.}.


The unusual flaring state of this otherwise steady source, the intensity of
the flare, and the experimental observations in $\gamma$-rays motivated this
search for neutrinos in IceCube in coincidence with the Crab flare of
Sep. 2010. The IceCube collaboration started a prompt analysis of the
then-running 79-string configuration.  The time window selected for this
analysis was the 10 days interval reported by ARGO-YBJ from September 17 to September 27, which contains the
Fermi flare window. An unbinned maximum likelihood (LLH) method described in
Ref.~\cite{Braun} has been applied to search for an excess of neutrinos in
coincidence with the enhanced $\gamma$-ray emission from the Crab.
The non observation of neutrinos would reinforce
pure electromagnetic emission scenarios and determine the level at which
hadronic phenomena superimposed on an electromagnetic scenario can be probed.

The IceCube Neutrino Observatory is a neutrino telescope installed in the
deep ice at the geographic South Pole. The final configuration comprises
5,160 photomultipliers (PMTs) \cite{PMT} along 86 strings instrumented
between 1.5-2.5 km in the ice. Its design is optimized for the detection of
high energy astrophysical neutrinos with energies above $\sim 100$ GeV. The
observation of cosmic neutrinos will be a direct proof of hadronic particle
acceleration and will reveal the origins of cosmic rays (CR) and the possible
connection to shock acceleration in Supernova Remnants (SNR), Active Galactic
Nuclei (AGN) or Gamma Ray Bursts (GRBs). The IceCube detector uses the
Antarctic ice as the detection volume where muon neutrino interactions
produce muons that induce Cherenkov light. The light propagates through the
transparent medium and can be collected by PMTs housed inside Digital Optical
Modules (DOMs). The DOMs are spherical, pressure resistant glass vessels each
containing a 25 cm diameter Hamamatsu photomultiplier and its associated
electronics. Eight densely instrumented strings equipped with higher quantum efficiency DOMs form, together with 12 adjacent IceCube strings, the DeepCore array that increases the sensitivity for low energy neutrinos down to about 10~GeV. Detector construction finished during the austral summer of
2010-11.

This paper describes in Sec.~\ref{sec:data} the data selection, the
comparison to simulation, and the detector effective area and angular
resolution for this search; in Sec.~\ref{sec:method} we summarize the
analysis method used; in Sec.~\ref{sec:results} the results for the flare
search are presented. Given the null result, upper limits are provided.  In
Sec.~\ref{sec:models} the time-integrated upper limits based on 1 year of
data of the 40-string configuration are presented to summarize what is the
impact of the IceCube most sensitive limit on existing neutrino production
models for the Crab. Conclusions are given in Sec.~\ref{sec:conc}.

\section{Data Selection and Comparison to Monte Carlo}
\label{sec:data}

The detection principle of IceCube is based on the charge and time
measurement of the Cherenkov photons induced by relativistic charged
particles passing through the ice sheet. The PMT signal is digitized with
dedicated electronics included in the DOMs \cite{WF}.  A DOM is triggered
when the PMT voltage crosses a discriminator threshold set at a voltage
corresponding to about 1/4 photoelectron. Various triggers are used in
IceCube. The results shown here are based on a simple multiplicity trigger
requiring that the sum of all triggered DOMs in a rolling time window of
$5\,\mu$s is above 8 (SMT8). The duration of the trigger is the amount of
time that this counter stays at or above 8 as the time window keeps
moving. Once the trigger condition is met, all local coincidence hits are
recorded in a readout window of $\pm 10\,\mu$s for the 40-string run and of
$^{+6}_{-4} $ $\mu$s (to reduce the noise rate) in the 79-string run.
IceCube triggers primarily on down-going muons at a rate of about 1.8~kHz in
the 79-string configuration.  Variation in the trigger rate determined by
atmospheric muons is about $\pm10\%$ due to seasonal changes \cite{Tilav}.
Seasonal variations in atmospheric neutrino rates are expected to be a
maximum of $\pm 4\%$ for neutrinos originating near the polar regions.  Near
the equator, atmospheric variations are much smaller and the variation in the
number of events is expected to be less than $\pm 0.5\%$
\citep{Ackermann:2005icrc}.

For searches of neutrino point sources in the northern sky, IceCube can use
the Earth as a shield to reduce the background of atmospheric muons and
detect up-going muons induced by
neutrinos. 
In the northern sky these searches are sensitive to neutrinos in the TeV-PeV
region.


In order to reconstruct muon tracks a LLH-based reconstruction is performed
at the South Pole (L1 filter) providing a first order background rejection of
poorly reconstructed events and a selection of high energy muons for the
southern sky. The data sent through the satellite to the North undergo
further processing that includes a broader range of more CPU consuming
reconstructions.  This offline processing also provides useful variables for
background rejection, measurements of the energy and of the angular
uncertainty, and selects about 35 Hz of the SMT8 data.  However, the offline
processing requires a fair amount of time to be finalized and is not suitable
for expedited analysis.  For the analysis of the Crab flare we used a
dedicated selection for target of opportunity programs
\cite{ToO}. 
This online event selection and reconstruction is called the online Level 2
filter and selects about 4 Hz of data. It provides a reduced data rate
(compared to the standard online data) because of stricter cuts than in the
offline filter. The loss of sensitivity of this stream of data is marginal
for $E^{-2}$ neutrino spectra.

\begin{figure}[htpb]  
\center
\includegraphics[scale=0.4]{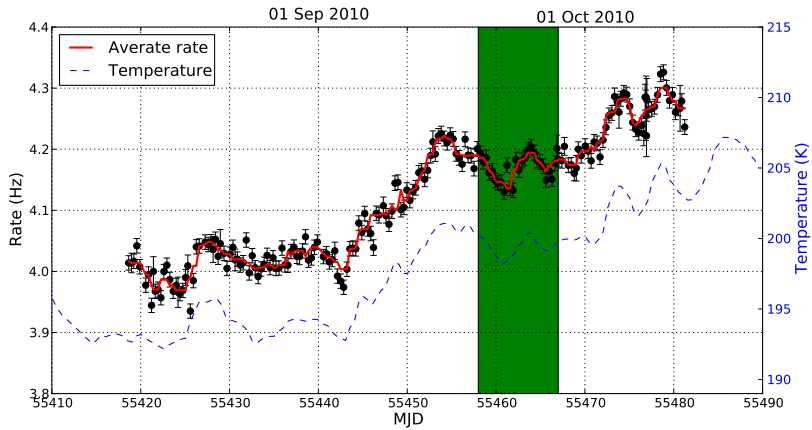}
\caption{\label{fig:RatesOnlineL2} Data rates per run of the online Level 2 filter of 79 IceCube strings in the time window of the Crab flare. The green bar indicates the flaring interval used in this analysis according to ARGO measurements~ \cite{ARGO}. The blue dotted line indicates the temperature in the middle stratosphere of the South Pole according to~\cite{Tilav}.}
\end{figure}

The online L2 filter performs a 8-fold iterative single photoelectron (SPE)
LLH fit for events with the number of DOMs triggered fewer than 300 and a 4-fold
iterative SPE fit otherwise. These SPE fits are seeded by a track obtained
using a single iteration LLH fit~\cite{track}. 
While the online Level 2 selects good quality tracks
and high energy muons from the northern sky, it is dominated by the
background of down-going atmospheric muons and therefore further cuts have to
be applied before performing neutrino source searches.  Experimental and
simulated data are processed and filtered in the same way.  The data used for
this search concern the period from 2010/08/10 to 2010/10/12.  In this period
the detector was running in a stable configuration. The total live time for
that period (considering deadtimes) is 60.9
days. Figure~\ref{fig:RatesOnlineL2} shows the data rate of each run included
during the selected time window as well as the South Pole atmospheric
temperature. As can be seen at this level, the rate is dominated by
down-going atmospheric muons, which display larger weather-dependent
variations than the final up-going neutrino events.

We have performed two dedicated selections starting from the online L2 filter that we describe below. 

\subsection{Straight Cuts Data Selection}

This dataset is obtained by requiring a good level of reconstruction and
ensuring degree level accuracy in the tracking errors to reject the
misreconstructed down-going atmospheric muons from the real up-going
atmospheric neutrino sample. The variables used are determined in the offline
data processing and have been used for the 40-string point-source analyses
in~\cite{jon} and \cite{mike}. The final cut level can be achieved by
applying the following series of cuts on a number of variables to obtain a
good agreement between data and the simulation of atmospheric neutrinos, with
a contamination of the order of 5\% of atmospheric muons, mainly muons from
two cosmic ray showers in coincidence in the same readout window. Having
these muons with different directions gives hit patterns that confuse the
reconstruction so that at times the result is a misreconstructed up-going
track. The cuts are:

\begin{equation}
N_{dir} \ge 5\;;\;L_{dir} > 200~m\;;\; \sigma_{cr} < 5^{\circ}\;;\; L_{red} \le \left\{\begin{array}{rl}  7.4  & \mbox{if  $L'_{red} \le$ 6.4} \\  8.0 & \mbox{otherwise.} \end{array} \right.
\end{equation}

where:
\begin{itemize}	
\item ${\bf N_{dir}}$: is the number photons detected within -15 and 75~ns
  with respect to the expected arrival time of unscattered photons from the
  reconstructed muon-track. Scattering of photons in the ice causes a loss of
  directional information and will delay them with respect to the unscattered
  expectation;
\item ${\bf L_{dir}}$: is the maximum distance in meters between direct
  photons projected along the best muon track solution;
\item ${\boldsymbol \sigma_{cr}}$: is the uncertainty on the reconstructed
  track direction given by the LLH-based track reconstruction estimated by a
  method based on the Cramer-Rao inequality \cite{cramer}; and
\item ${\bf L_{red}}$ and ${\bf L'_{red}}$: are the standard reduced and
  modified LLH values respectively. The reduced LLH is defined as the $-\log_{10}$ of the
  LLH value of the track reconstruction divided by the number of degrees of
  freedom. The number of degrees of freedom is the number of hit DOMs minus
  five fit parameters, two angles and three coordinates of a reference point
  along the track. It was found by comparing background rejection efficiency
  to signal selection efficiency that a good variable for rejection of
  background for low energy events is the number of hit DOMs minus an
  effective number of degrees of freedom of 2.5.
\end{itemize}

An additional cut to select events in the direction of the Crab
($\Theta_{\rm{Crab}}=122^\circ$ at the South Pole) has also been applied:
$\Theta_{\rm{Crab}} - 10^{\circ} < \theta_{rec} < \Theta_{\rm{Crab}} +
10^{\circ}$, where $\theta_{rec}$ is the reconstructed zenith angle of the
muon track. No further selection in right ascension has been applied. In
Tab.~\ref{tab: rates} the selected number of events and the expected number
of atmospheric neutrinos and muons are given. The final number of events
selected for the 10 day window of the flare is 354.

\subsection{BDT Data Selection}

The second dataset is obtained by using a multi-variate learning machine. In
particular this data selection is based on the knowledge and experience from
previous analyses looking for solar Weakly Interactive Massive Particles
(WIMPs) with the IceCube detector \cite{wimp_ic22}. During the austral winter
the Sun is below the horizon at the South Pole and its maximum declination is
equal to the obliquity of the ecliptic, $23.4^{\circ}$.  Since the Crab
Nebula lies fairly close to the ecliptic plane, the strategies and cuts that
are optimized for this specific direction can be applied for the Crab
direction.

Starting with the online L2 filtered data selection, as described above, a
number of additional cuts were applied. The hereby selected events fulfill
criteria of horizontal tracks passing the detector, to further minimize
vertical tracks associated with background events. Additionally, the cuts
were chosen to reduce the tails of distributions of the background into the
signal region:

\begin{equation}
z_{travel} > -10~m\;;\;\sigma_{COGz} < 170~m \;;\; \sigma_{cr} < 10^{\circ}\;;\;\rho_{av}  < 150~m\;;\;t_{accu} < 3000~ns
\end{equation}

where:
\begin{itemize}
\item ${\boldsymbol z_{travel}}$: measures the difference in the $z$
  positions of the center of gravity (COG) of the hits at the beginning of an
  event (first 1/4 of the hits in time) and the COG at the end of the event
  (last 3/4 of the hits in
  time); 
\item ${\boldsymbol \sigma_{COG_{z}}}$: is the uncertainty in meters of the $z$-coordinate of the COG;
\item  ${\boldsymbol \rho_{ave}}$: is the mean minimal distance between the LLH track and the hit DOMs; and
\item ${\boldsymbol t_{accum}}$: is the accumulation time, defined as the
  time until 75\% of the total charge develops in ns.
\end{itemize}

Boosted Decision Trees (BDTs) \cite{BDT}, multi-variate learning machines,
were used in the final analysis step to classify events as signal-like or
back\-ground-like. Eleven event observables, split in two sets of 5 and 6
each, were obtained by choosing parameters with low correlation in background
(correlation coefficient $|c| < 0.5$), but high discriminating power between
signal and background.  The selected observables include $N_{dir}$,
$L_{dir}$, $\sigma_{cr}$ and $L'_{red}$ as described within the straight cuts
data selection in Sec.2.1 and $z_{travel}$ from above. Additionally,
observables specifying the geometry, the time evolution of the hit pattern,
the quality and consistency of the various track reconstructions that is
defined through the opening angle between the line-fit and the LLH tracks,
and the number of hit strings are used.  Training was done with simulated
signal events for a soft neutrino spectrum of $E^{-3}$ that also well
represents the case of an $E^{-2}$ spectrum with a TeV cut-off. A set of
off-time real data, not used in the flare analysis, was used for training as
background.  The final sample is defined by a cut on the combined output
(score) of the two BDTs. As in the case of the straight cuts sample, an
additional requirement of reconstructed zenith tracks within $\pm 10^\circ$
from the Crab has been applied. In Tab.~\ref{tab: rates} the selected number
of events and the expected number of atmospheric neutrinos and muons are
given. The final number of events selected for the 10 day window of the flare
is 660 events in the northern sky.

\begin{table}[h]
\begin{center}
\hspace{-0.0cm}
\begin{tabular}{ c|c|c|c|c } \toprule
\multirow{2}{*}{Cut Level} & Data  rate& Atm. $\mu$ rate & Atm. $\nu_{\mu}$ rate & E$^{-2}$ Eff.  \\
& (Hz) &  (Hz) & (Hz) & (\%) \\
\midrule
Trigger & 1,800 & 1,800 & 2.59 $\times 10^{-2}$ &  - \\
Online Level 2 &  4.03 &  3.11 &  7.2 $\times 10^{-3}$ & 100\\
Straight Cuts &  $4.6 \times 10^{-4}$ & $\sim 0$ & 4.8 $\times 10^{-4}$ & 55\\
BDT &  $8.4 \times 10^{-4}$ & $\sim 0$ & $8.2 \times 10^{-4}$ & 61\\ 
\bottomrule
\end{tabular}
\vspace{0.5 cm}
\caption{Data, atmospheric muon, and neutrino expected background rates for different cut progression. The signal efficiency for an $E^{-2}$ neutrino spectrum assuming an emission $\pm 10^\circ$ around the Crab with respect to the online Level 2 is also shown.}
\label{tab: rates}
\end{center}
\end{table}

\subsection{Comparison Data-Monte Carlo and Detector Performance}

The simulation of atmospheric and signal neutrinos that is used for
determining the selection efficiency, the performance of the detector and to
calculate upper limits is based on the neutrino generator ANIS
\cite{Gazizov:2004va} and the deep inelastic neutrino-nucleon cross sections
with CTEQ5 parton distribution functions \cite{Lai:1999wy}.  Neutrino
simulation can be weighted for different fluxes, accounting for the
probability of each event to occur.  In this way, the same simulation sample
can be used to represent atmospheric neutrino models such as Bartol
\cite{Barr:2004br} and Honda \cite{Honda:2006qj} neutrino fluxes from pion
and kaon decays (conventional flux) and a variety of models for the charm
component (prompt flux) \cite{Martin:2003us,Enberg:2008te}.  Muons from CR
air showers were simulated with CORSIKA \cite{Heck:1998vt} with the SIBYLL
hadronic interaction models \cite{Ahn:2009wx}. An October polar atmosphere, an
average case over the year, is used for the CORSIKA simulation. Seasonal
variations are therefore to be expected less than $\pm 10\%$ in event rates
\citep{Tilav}. Muon propagation through the Earth and ice are done using MMC
\citep{Chirkin:2004hz}.  This simulation is used to verify the level of
agreement of data and MC from trigger level to Level 1 and to understand the level
of contamination at final cut level.  For the optical properties of the ice
we used a model obtained from calibrations using the LEDs in the DOMs called
flashers \cite{Dima}. This model produces a better agreement between data and
MC than the model previously used
\cite{OpticalProperties_JGeophysRes_2006}. The simulation propagates the
photon signal to each DOM using light tracking software described in
\cite{Lundberg:2007mf}. The simulation of the DOMs includes their angular
acceptance and electronics.  The systematic errors on the simulation of the
signal used to produce the upper limits have been evaluated and presented in
Sec. 6 of Ref. \cite{jon} describing the 40-string time-integrated point
source search. The main uncertainties on the limits for an $E^{-2}$ signal of
muon neutrinos come from photon propagation, absolute DOM efficiency, and
uncertainties in the Earth density profile and muon energy loss, accounting
for a total of 16\%.

\begin{figure}[htpb]  
\hspace{-1cm}
	\begin{tabular}{c c}
		\includegraphics[scale=0.4]{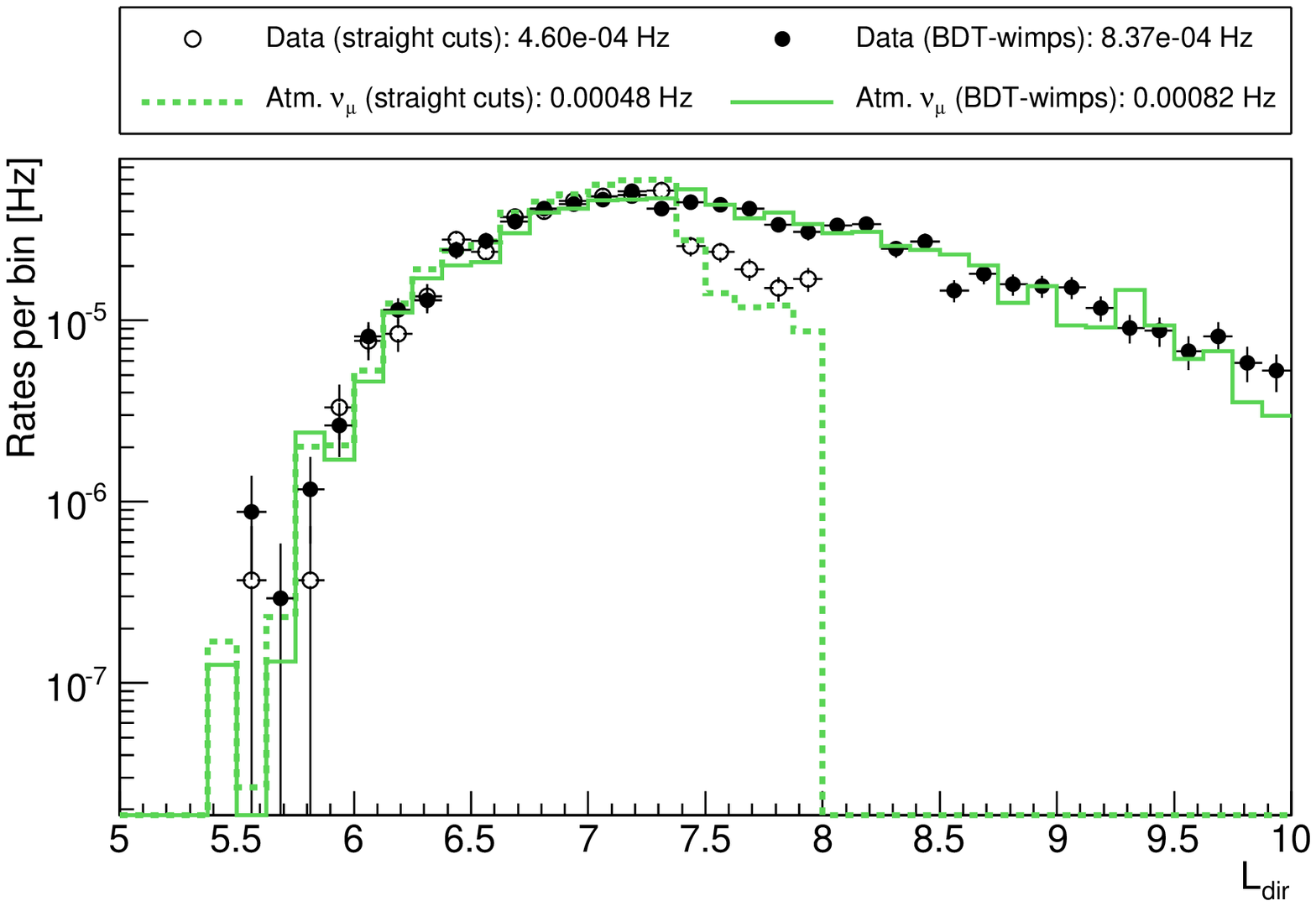} \hspace{-0.5cm}
		\includegraphics[scale=0.4]{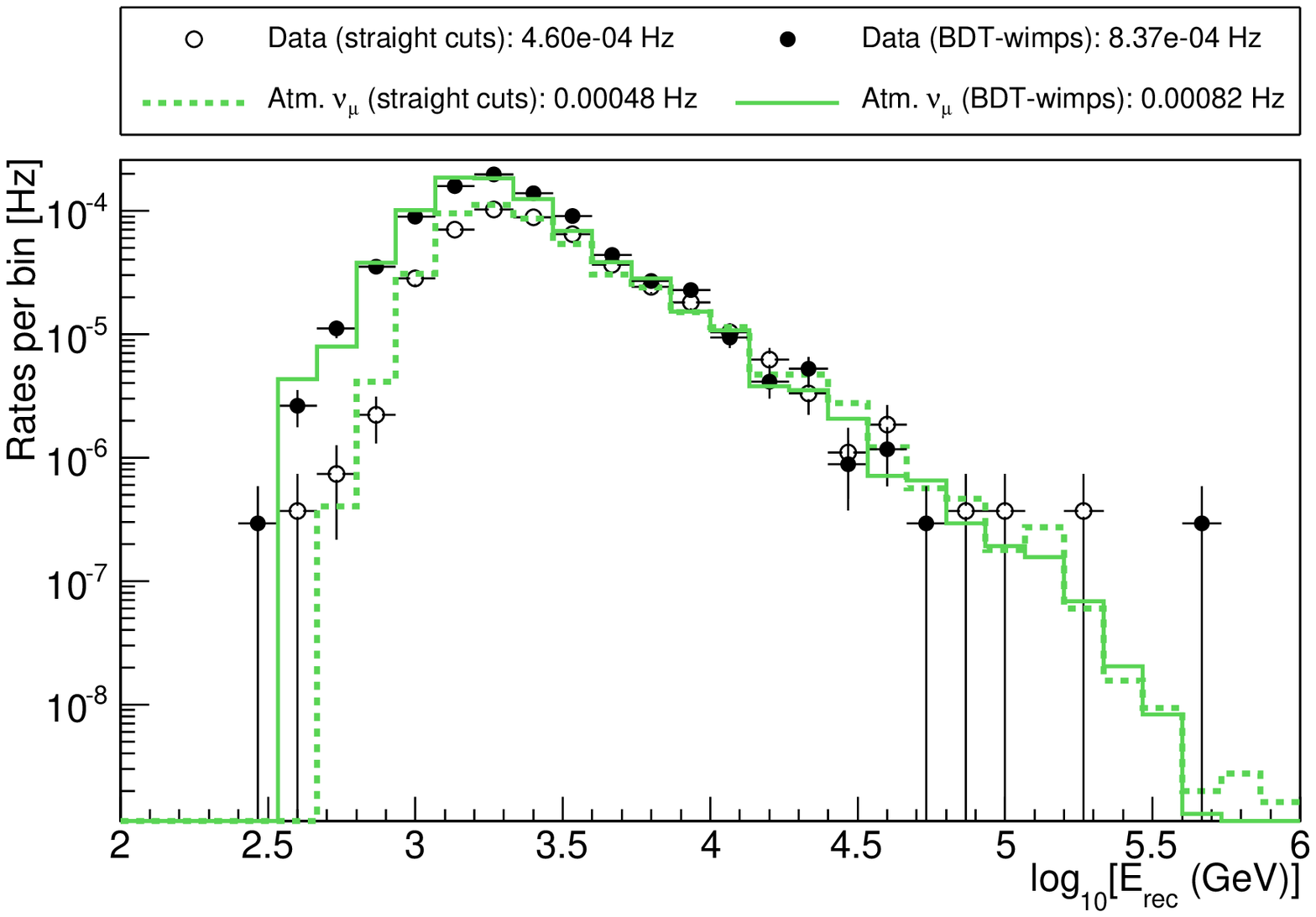}\\
		\includegraphics[scale=0.4]{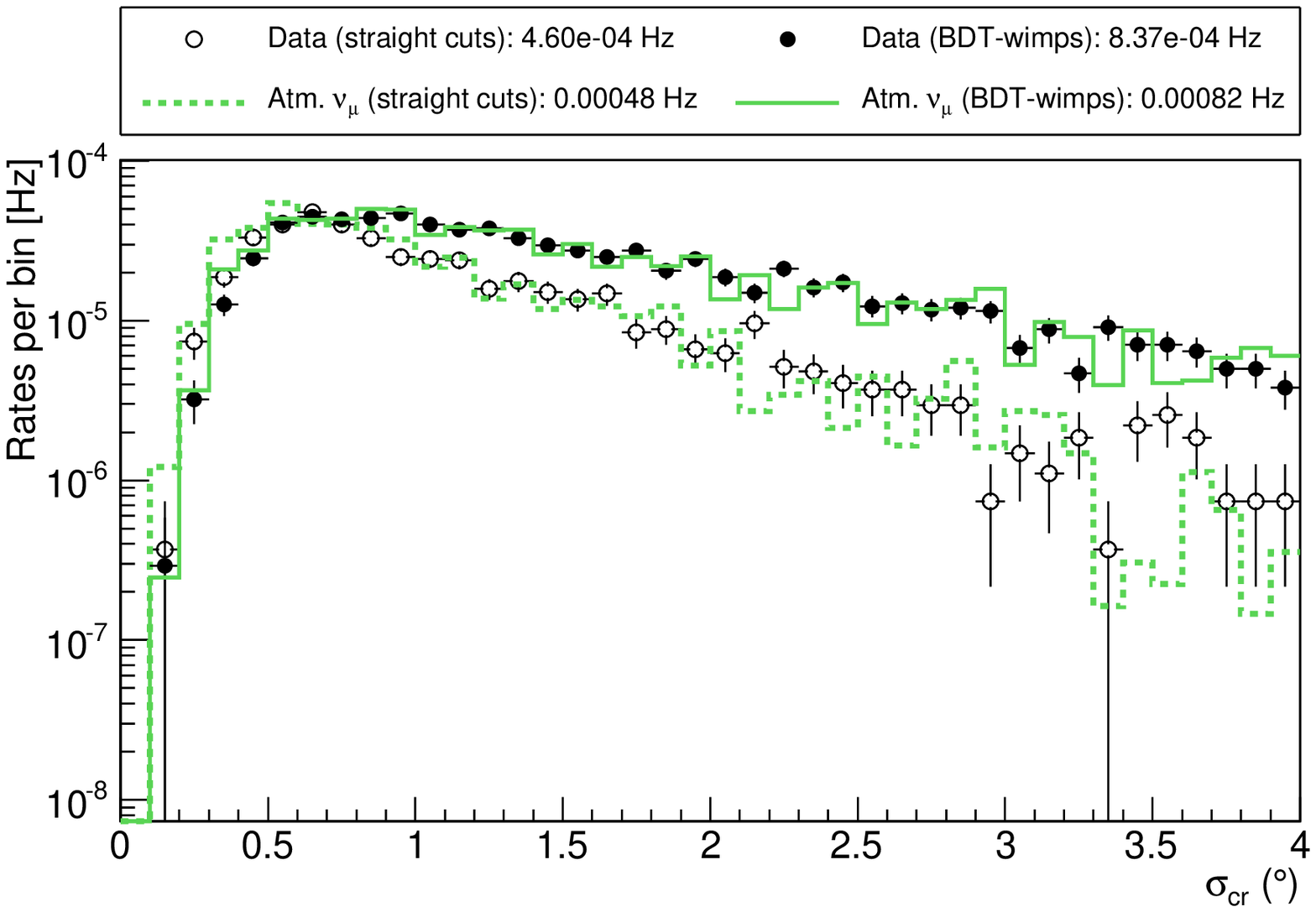} \hspace{-0.5cm}
		\includegraphics[scale=0.4]{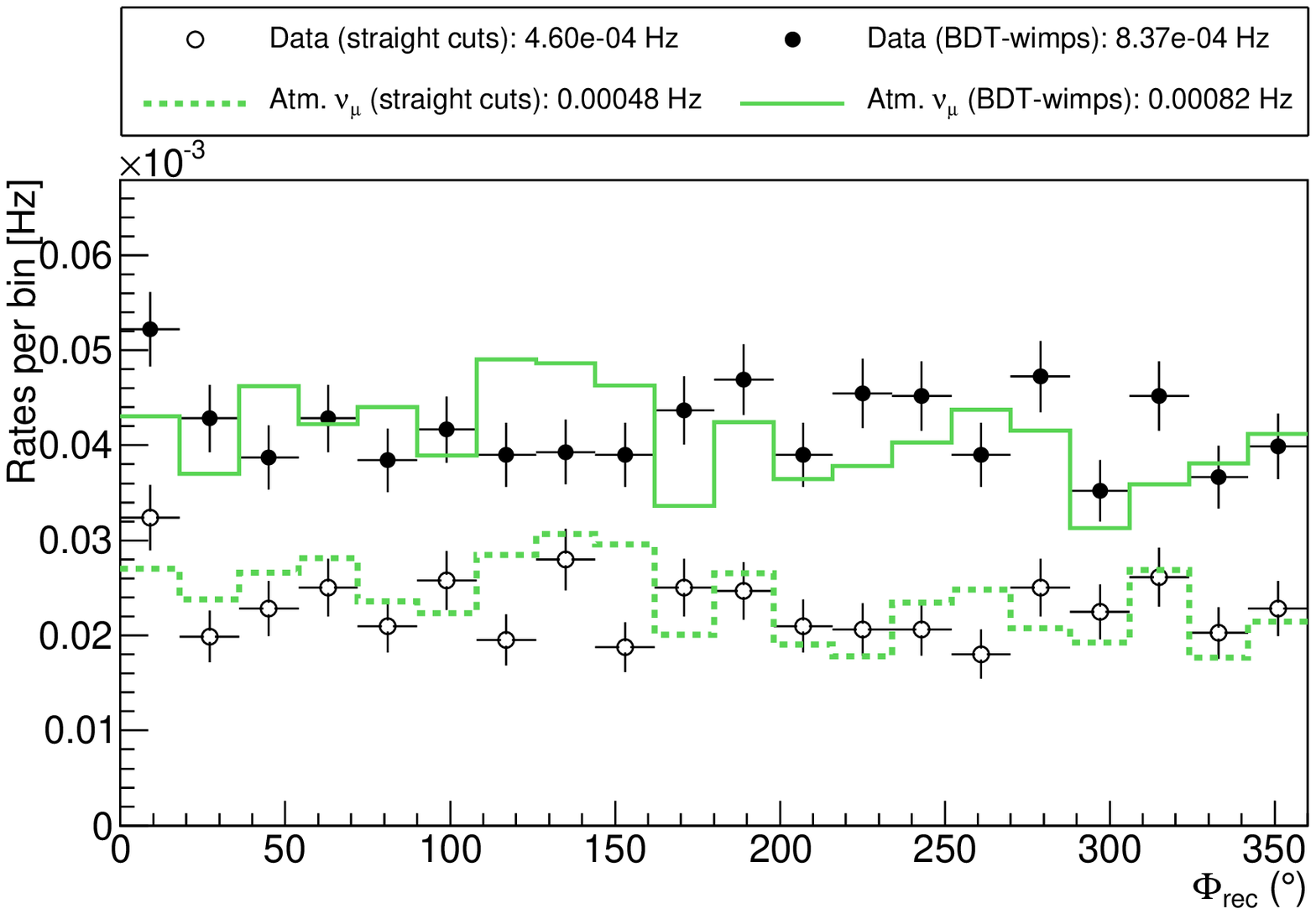} 
	\end{tabular}
\caption{\label{fig:dataMC} The top-left plot shows the  reduced log-likelihood ($L_{red}$), as defined in section~\ref{sec:data}, distribution for both data (dots) and atmospheric neutrino simulation (green lines) for the two data samples. The distribution of the reconstructed energy is shown on the top-right plot.  The estimated angular error given by the track reconstruction algorithm using the Cramer-Rao upper bound is shown on the bottom-left plot while the bottom-right shows the azimuth distribution of the final data samples.}
\end{figure}

\begin{figure}[htpb]  
\hspace{-1cm}
\begin{tabular}{c c}
\includegraphics[scale=0.4]{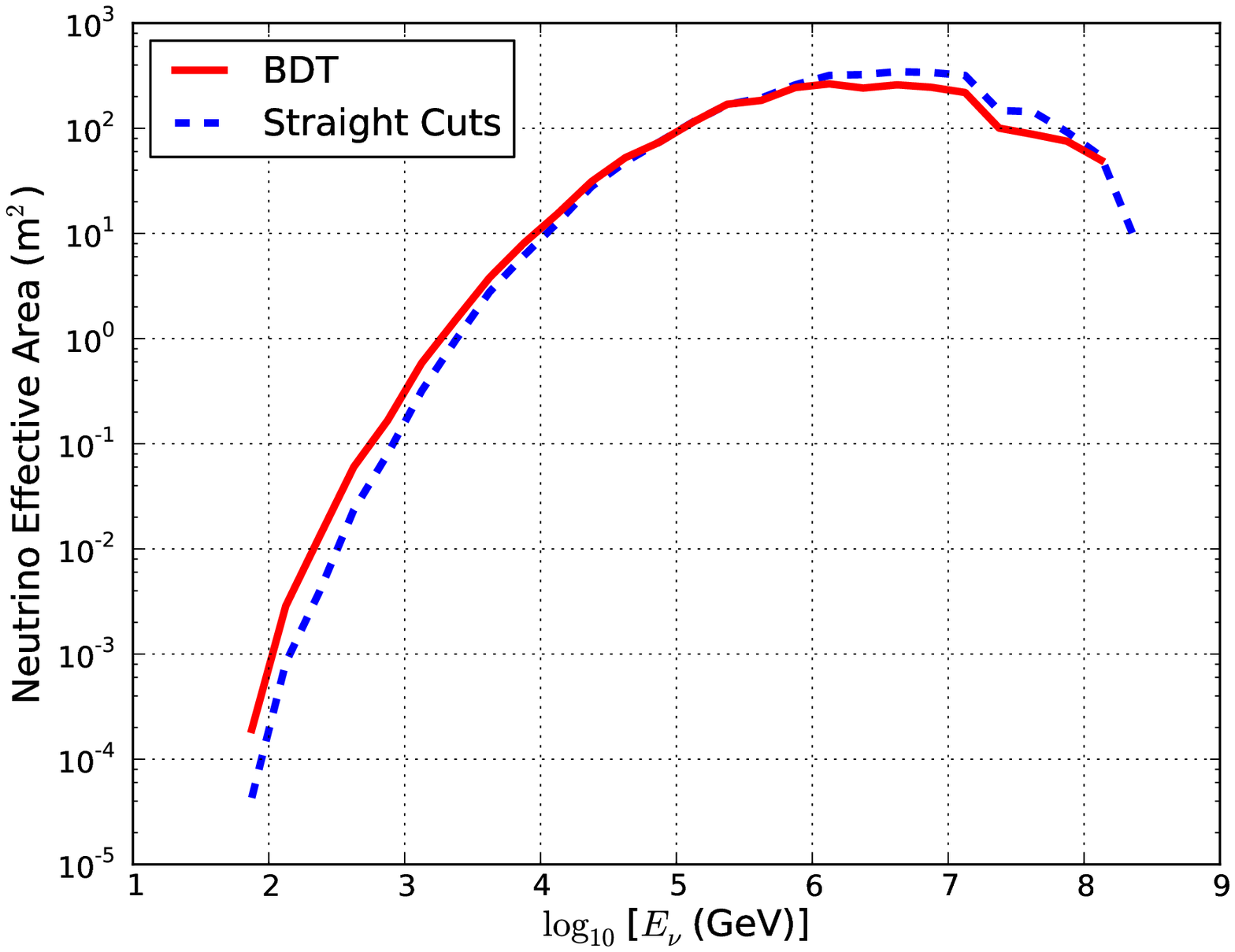}
\hspace{-1cm}
\includegraphics[scale=0.4]{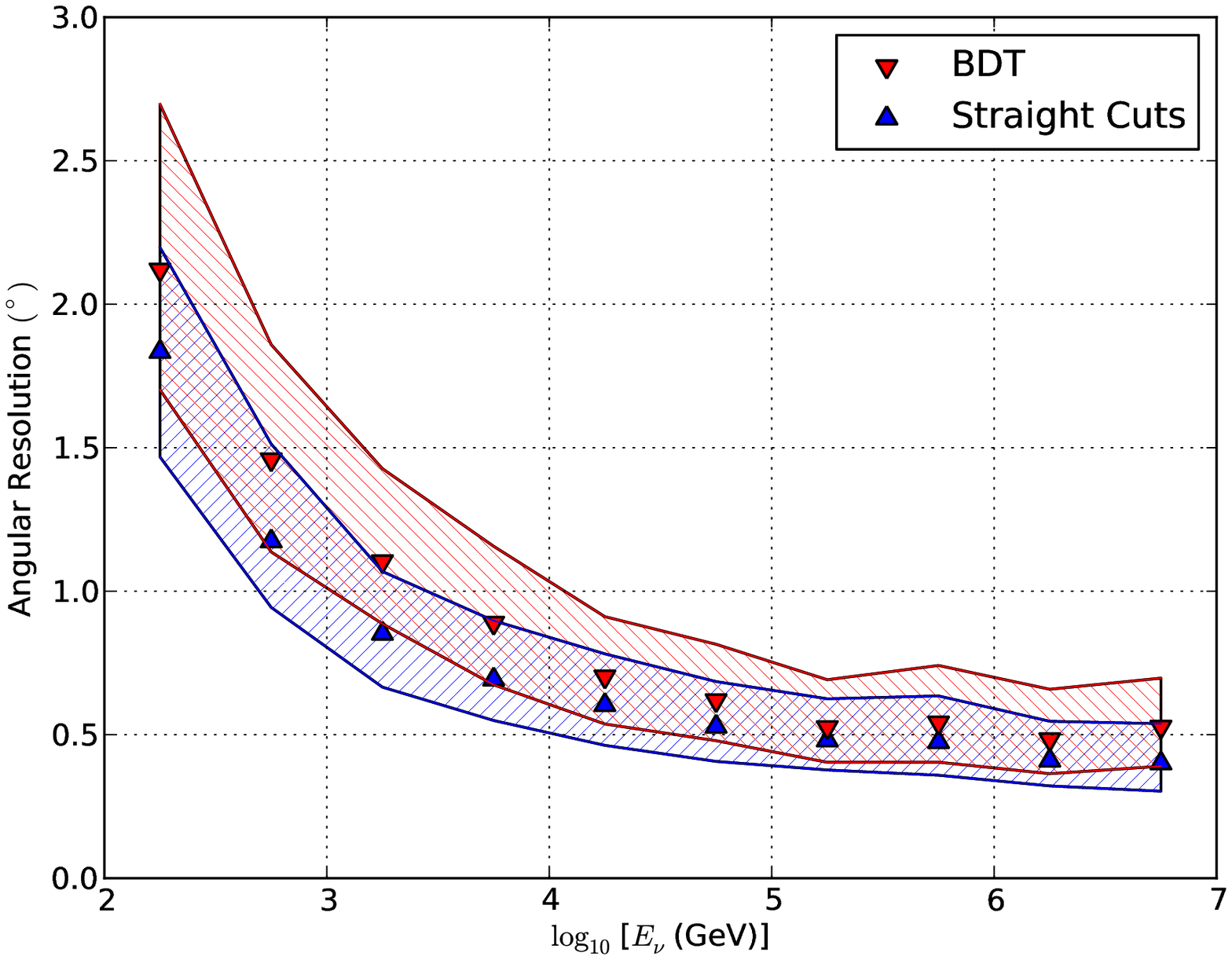} 
\end{tabular}
\caption{\label{fig:AeffandResol} Left: Muon neutrino effective area for the two
  final data samples in a zenith bin of $\pm 10^\circ$ from the direction of
  the Crab Nebula. Right: Angular resolution defined as the median of the
  point spread function as a function of the neutrino energy for the two data
  samples. The shaded areas represent a $\pm 10$\% area of the point spread
  function.}
\end{figure}

Figure~\ref{fig:dataMC} shows the data and simulation comparison for some
variables at the final cut level for the two data samples.  As can be seen,
the BDT sample increases the overall rate by allowing more low quality
reconstructed events (high $L_{red}$) than the straight cut sample. This is
translated into a higher neutrino effective area at low energies but also a
worse angular resolution as can be seen in figure~\ref{fig:AeffandResol}.

\section{Likelihood analysis}
\label{sec:method}

The method used for this analysis is an unbinned likelihood method
\cite{Braun}. This method looks for a localized statistically significant
excess of neutrinos above the background in the direction of the Crab in
coincidence with the flare. The same analysis technique has already been
applied to AGN flare searches in IceCube~\cite{mike}. The method uses both
the reconstructed direction of the events as well as an energy proxy, the
reconstructed visible muon energy, to discriminate any possible signal from
background during the time interval of the flare. We consider the largest reported time window of 10 days by
ARGO-YBJ. The applied method describes the data as a two component mixture of
signal and background.  For a data set with N total events the probability
density of the $i^{th}$ event is given by:

\begin{equation}
\frac{n_s}{N}\mathcal{S}_i + ( 1-\frac{n_s}{N} )\mathcal{B}_i,
\label{eq:pdf}
\end{equation}

\noindent where $\mathcal{S}_{i}$ is the density distribution for the signal
hypothesis and $\mathcal{B}_{i}$ for background. The parameter $n_{s}$ is the
number of signal events and one of the free parameters of the likelihood
maximization together with the spectral index, $\gamma$, of the signal
spectrum distribution.  The likelihood of the data is the product of all
event probability densities:
\begin{equation}
\mathcal{L}(n_s, \gamma) = \prod_{i=1}^{N} \Big[\frac{n_s}{N}\mathcal{S}_i + (1 - \frac{n_s}{N})\mathcal{B}_i\Big].
\label{eq:lh}
\end{equation}
The likelihood is then maximized with respect to $n_s$ and $\gamma$, giving
the best fit values $\hat{n}_s$ and $\hat{\gamma}$.  The null hypothesis is
given by $n_s=0$ ($\gamma$ has no meaning when no signal is present).  The
likelihood ratio test-statistic is defined as:
\begin{equation}
TS = -2 \log\Big[\frac{\mathcal{L}(n_s=0)}{\mathcal{L}(\hat{n}_s,\hat{\gamma}_s)}\Big].
\label{eq:ts}
\end{equation}
The background probability distribution function, or pdf, $\mathcal{B}_i$, is given by:

\begin{equation}
\mathcal{B}_{i} = \mathcal{B}^{space}_{i}(\theta_{i},\phi_{i}) \mathcal{B}^{energy}_{i}(E_{i},\theta_{i}) \mathcal{B}^{time}_{i}(t_{i},\theta_{i}),
\label{llh_bg_time1}
\end{equation}
and is computed using the distribution of data itself.  The spatial term
$\mathcal{B}^{space}_{i}(\theta_{i},\phi_{i})$ is the event density per unit
solid angle as a function of the local coordinates. The energy probability,
$\mathcal{B}^{energy}_i(E_{i},\theta_{i})$, is determined from the energy
proxy distribution of data as a function of the cosine of the zenith angle, $\theta_{i}$.  This energy proxy, described in detail in~\cite{jon}, uses
the density of photons along the muon track due to stochastic energy losses
of pair production, bremsstrahlung and photonuclear interactions which
dominate over ionization losses for muons above 1 TeV.  The time probability
$\mathcal{B}^{time}_{i}(t_{i},\theta_{i})$ of the background can be taken to
be flat for this case of a 10 day time interval ignoring the seasonal
modulations.

The signal pdf $S_i$ is given by:
\begin{equation}
  \mathcal{S}_{i}=\mathcal{S}^{space}_i(\mid \vec{x}_i-\vec{x}_{s} \mid, \sigma_{i})\mathcal{S}^{energy}_i(E_{i},\theta_{i},\gamma_{s})\mathcal{S}^{time}_{i}(t_{i}),
\label{llh_signal_time1}
\end{equation}
where $\mathcal{S}^{space}_i$ depends on the angular uncertainty of the event
$\sigma_{i}$ and the angular difference between the event position
$\vec{x}_i$ from the source position $\vec{x}_{s}$.  The density function
$\mathcal{S}^{energy}_i$ is a function of the reconstructed energy proxy
$E_{i}$, and the spectrum $\gamma_{s}$ is calculated from an energy
distribution of simulated signal in a zenith band that contains the
source. The signal time probability, $\mathcal{S}^{time}_{i}$, depends on the
particular signal hypothesis. In this analysis we adopt a simple cut in time
between $t_{min}$ and $t_{max}$, which can be expressed as:
\begin{equation}
\mathcal{S}^{time}_i = \frac{H(t_{max}- t_i) \times H(t_i -t_{min})}{t_{max}-t_{min}},
\end{equation}
where $t_i$ is the arrival time of the event, $t_{max}$ and $t_{min}$ are the upper and lower bounds of the time window defining the flare, and H is the Heavyside step function. 

The significance of the result is evaluated by comparing the test-statistic
with a distribution obtained by performing the same analysis over a set of
background-only scrambled data sets.  The fraction of trials above the
test-statistic value obtained from data is referred to as the $p$-value, with
smaller $p$-values indicating that the background-only (i.e. null) hypothesis
is increasingly disfavored compared to the signal-plus-background hypothesis
as a description of the data. This leads to the definition of the discovery
potential: the average number of signal events required to achieve a
$p$-value less than 2.87$\times 10^{-7}$ (one-sided 5$\sigma$) in 50\% of
trials. Similarly, the sensitivity is defined as the average signal required
to obtain, in 90\% of trials, a test-statistic greater than the median
test-statistic of background-only scrambled samples.

\section{Results}
\label{sec:results}

\begin{figure}[htpb]  
\hspace{-1cm}
\begin{tabular}{c c}
\includegraphics[scale=0.4]{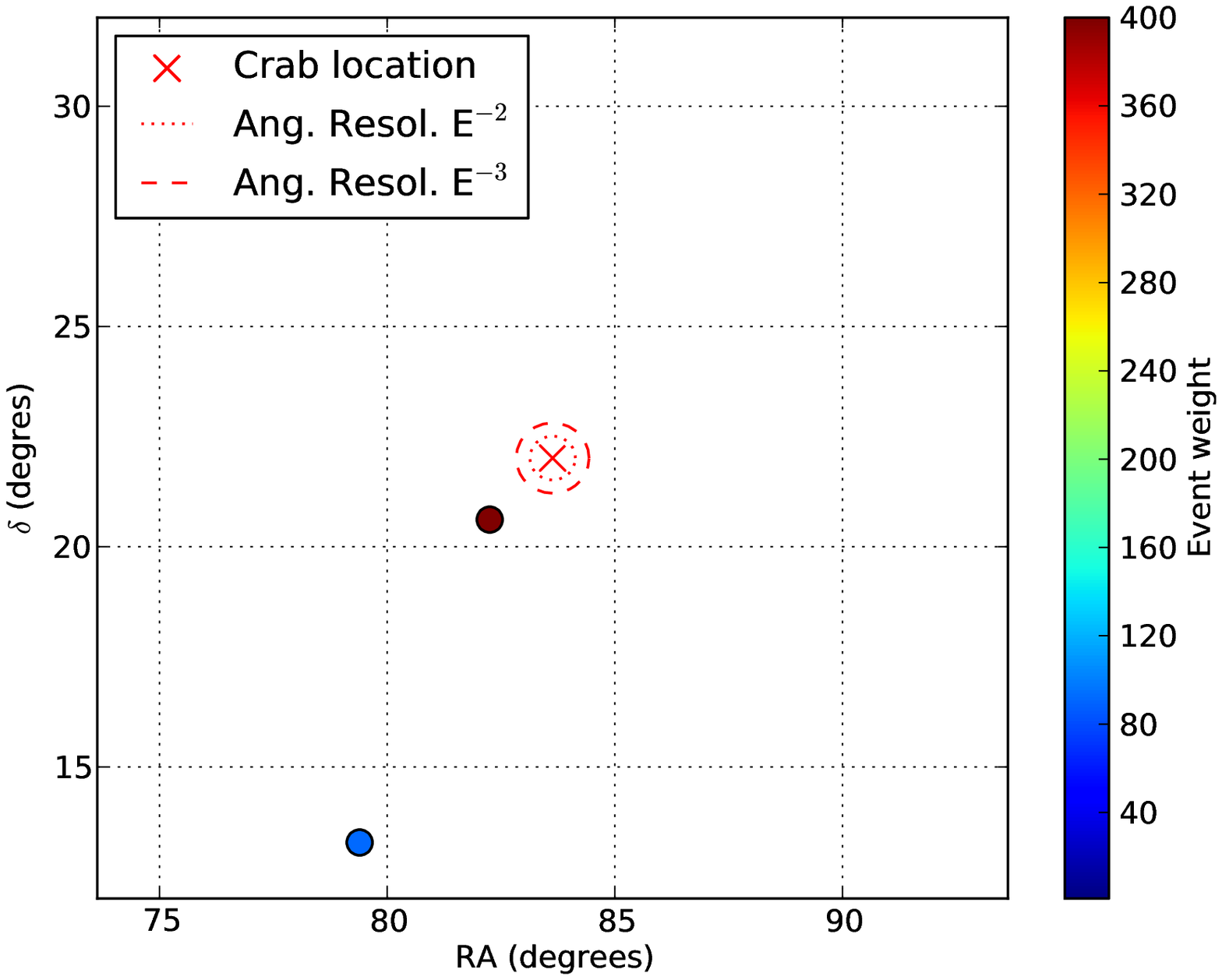}
\hspace{-1cm}
\includegraphics[scale=0.4]{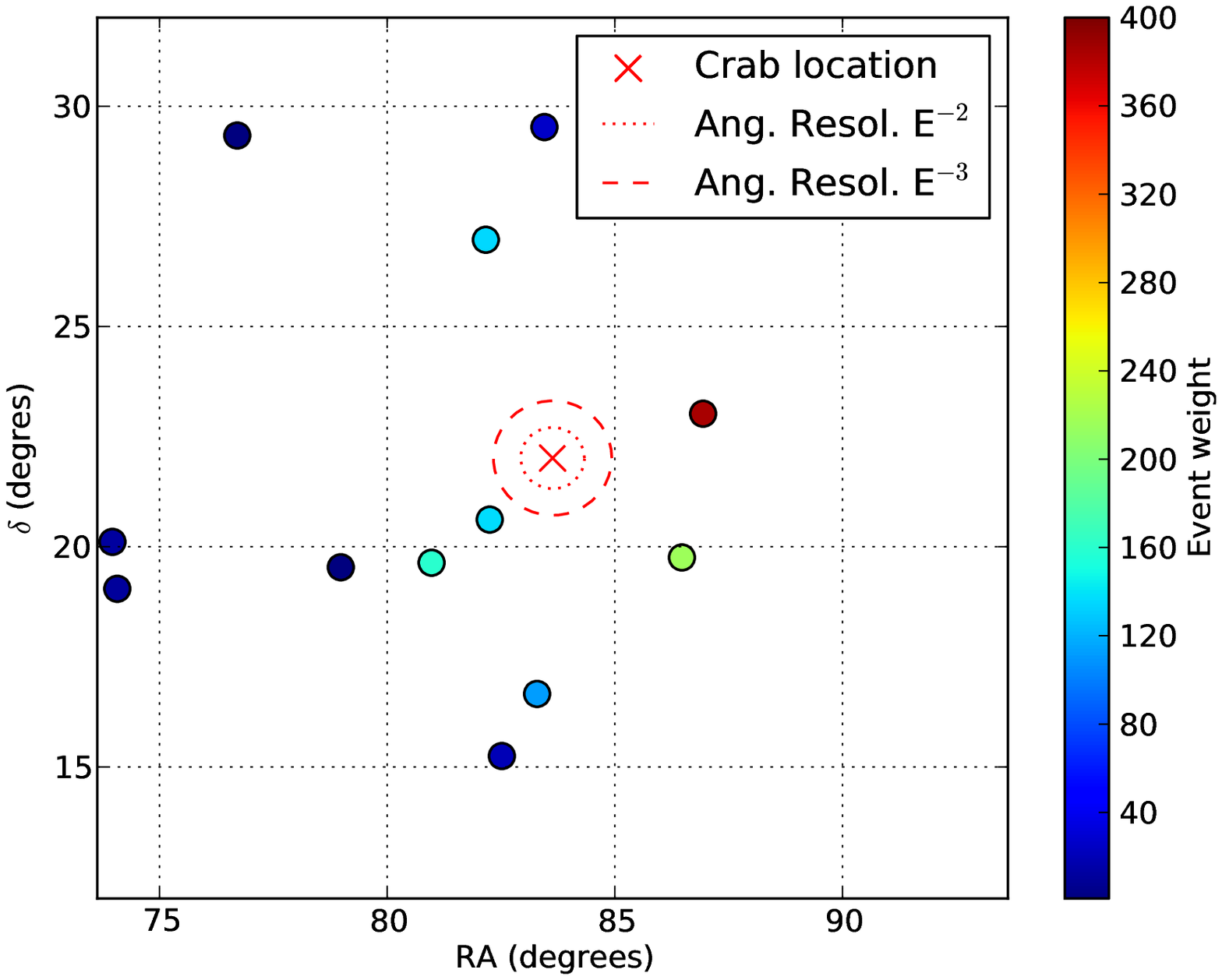} 
\end{tabular}
\caption{\label{fig: scatter} Left: Event distribution for the straight cuts for events with $\frac{\mathcal{S}_{i}}{\mathcal{B}_{i}} > 1$. The color code represents the event signal-over-background ratio. Right: Same distribution for the BDT sample.}
\end{figure}

The method described in section~\ref{sec:method} has been applied to both
data samples, the one obtained with straight cuts and the one obtained using
the BDTs. In both cases the best fit resulted in $n_s = 0$ (i.e. an under-fluctuation). Figure~\ref{fig: scatter} shows the event distribution for those events with a $\frac{\mathcal{S}_{i}}{\mathcal{B}_{i}} > 1$, that is, only events inside the flare window that contribute to the likelihood.

As can be seen, due to its higher neutrino efficiency at energies below 10 TeV the BDT sample has more atmospheric neutrino events. Since the background estimation depends on the sample,
the signal-to-background ratios are different for the same events in the two
samples. The highest event weight comes from the straight cuts sample.

Table~\ref{tab: ul} shows the upper limits set by both data samples for
different neutrino spectra. Each upper limit is shown both in terms of number
of signal events that can be rejected at 90\% CL, $n_{s}^{90\%}$, and the
flux limit on muon neutrinos, $\Phi^{90\%}_{\nu_{\mu}}$, for a 9.28 day
interval in units of cm$^{-2}$s$^{-1}$TeV$^{-1}$, i.e. $\frac{dN^{90\%}}{dE}
= \Phi^{90\%}_{\nu_{\mu}} \left( \frac{E}{\rm{TeV}}\right)^{-\gamma}$.


\begin{table}[h]
\begin{scriptsize}
\begin{center}
\hspace{-0.0cm}
\begin{tabular}{ l l | c c c c c | c c c c c} 
\toprule
\multicolumn{2}{c|}{Spectrum} &
\multicolumn{5}{c|}{Straight Cuts sample} &
\multicolumn{5}{c}{BDT sample} \\ 
\multirow{2}{*}{$E^{-\gamma}$} & {E$_{cutoff}$}   & \multirow{2}{*}{$n_{s}^{90\%}$} & \multirow{2}{*}{$\Phi^{90\%}_{\nu_{\mu}}$}& E$_{min}$ & E$_{max}$ & \multirow{2}{*}{$\mathcal{L}^{90\%}_{\nu_{\mu}}$}  & \multirow{2}{*}{$n_{s}^{90\%}$} & \multirow{2}{*}{$\Phi^{90\%}_{\nu_{\mu}}$} & E$_{min}$ & E$_{max}$& \multirow{2}{*}{$\mathcal{L}^{90\%}_{\nu_{\mu}}$}  \\
& (TeV) &  & & (GeV) & (GeV) &  & &  & (GeV) &
(GeV) &  \\
\midrule
$E^{-2}$ & -  & 2.15 & $4.84$ &10$^{3.3}$ & 10$^{5.7}$& 1.78 & 2.35 & $4.80$ & $10^{3.1}$ & $10^{5.9}$ & 2.02 \\
$E^{-2.7}$ &- & 2.41 & $32.6$ &10$^{2.6}$ & 10$^{4.9}$& 6.0 & 2.90 & $26.3$ & $10^{2.3}$ & $10^{4.7}$ & 6.77\\
$E^{-2}$ &1 & 2.80 & $309$ & 10$^{2.4}$ & 10$^{3.5}$ & 21.2 & 3.50 & $191$ & $10^{2.3}$&$10^{3.5}$&15.5\\
$E^{-2}$ &100  &  2.25 & $8.59$ & 10$^{3.2}$ &10$^{5.0}$& 1.98 & 2.51 & $8.06$&$10^{2.9}$&$10^{4.9}$&2.05\\
$E^{-2}$ &1000 &  2.20 & $5.52$&  10$^{3.3}$ & 10$^{5.6}$ &1.79 & 2.34 & $5.31$&$10^{3.0}$&$10^{5.5}$&1.86\\

\bottomrule

\hline
\end{tabular}
\vspace{0.5 cm}
\caption{Upper limits of the Crab Sep. 2010 flare using Neyman for both samples and
  different neutrino spectra including those with an exponential energy
  cut-off expressed as $E^{-\gamma} \exp\left({-E/E_{cutoff}}\right)$ where E$_{cutoff}$ is the energy cut-off. The
  number $n_{s}^{90\%}$ is the limit in terms of number of signal events for
  a 90\% confidence level and $\Phi^{90\%}_{\nu_{\mu}}$ is the flux upper
  limit in units of 10$^{-11}$cm$^{-2}$s$^{-1}$TeV$^{-1}$ for a 9.28 days flaring
  interval. The resulting neutrino luminosity limit, $\mathcal{L}^{90\%}_{\nu_{\mu}}$, is given
  in units of 10$^{35}$ erg s$^{-1}$ and it was calculated by integrating
  $dN^{90\%}/dE\times E$ over the energy range from $E_{min}$ to $E_{max}$ to
  contain 90\% signal of the spectrum and multiply by $4\pi d^{2}$ where $d$ is the distance to the Crab Nebula ($d = 1850$ pc).}
\label{tab: ul}
\end{center}
\end{scriptsize}
\end{table}

The analysis described and the results given in Tab.~\ref{tab: ul} rely on
the fact that background simulation can be performed by scrambling the right
ascension in real data (even if a signal is present in the data sample the
scrambling will dilute it over the background). This method of estimating the
background gives robust $p$-values in terms of systematic
uncertainties. Systematic uncertainties only affect the estimate of the
signal flux from the source and the upper limits. The systematic
uncertainties on the expected flux come from photon propagation in ice,
absolute DOM sensitivity ($\pm 8\%$), and uncertainties in the Earth density
profile as well as muon energy loss. The main uncertainty however is the
modeling of Antarctic ice and its effect on the photon propagation. In
IceCube different ice models have been devised. The variation in the upper
limits depending on the photon propagation model used are within $<
10$\%. Overall the uncertainty on upper limits is $16\%$.

\section{Impact of IceCube time-integrated limits on models from the Crab}
\label{sec:models}

\begin{figure}[htpb]  
 \center
\includegraphics[scale=0.5]{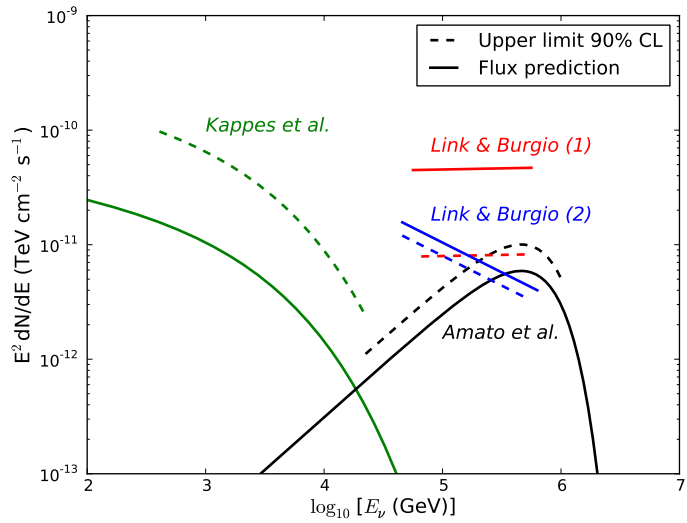}
\caption{\label{fig: models} Predicted fluxes and upper limits based on the IceCube 40 string configuration on several models from the Crab. Solid lines indicate the predicted flux and dotted lines the corresponding upper limit for a 90\% CL. The green lines are the predicted flux and corresponding upper limit based on the model proposed in~\cite{kappes}. The red and blue lines correspond to the model in~\cite{Burgio} for the cases of linear (1) and quadratic (2) proton acceleration. The black line represents the estimated flux for the most optimistic model proposed in~\cite{Blasi} based on resonant cyclotron absorption model and its corresponding upper limit.}
\end{figure}

The main goal of the IceCube telescope is the search for cosmic neutrino
signals that might explain the astrophysical phenomena that give rise to the
cosmic ray emission. In the absence of detection, constraining models can
also provide insights about the nature of these phenomena. The best available
neutrino flux limits for the Crab are based on the time-integrated analysis performed during the 375.5 d period
corresponding to the 40-string configuration of IceCube. We discuss
here the impact of these limits on different models of neutrino
emission from the Crab. Figure~\ref{fig: models} summarizes a number of different predicted
fluxes described in the introduction of this paper and where the 40-string
configuration limits stand~\cite{jon}. Upper limits are defined as the 90\%
confidence level (CL) using the method from Feldman \&
Cousins~\cite{FeldmanCousins}. The green line (solid) corresponds to the flux
predicted in~\cite{kappes} based on the $\gamma$-ray spectrum measured by
H.E.S.S. and the corresponding upper limit (dashed). The black line
represents the estimated flux based on the resonant cyclotron absorption
model proposed in~\cite{Blasi} for the case of a wind Lorentz factor of
$\Gamma = 10^{7}$ and the most optimistic case of the effective target
density.  The red and blue lines represent the two predicted fluxes according
to~\cite{Burgio} for the cases of linear and quadratic proton acceleration
respectively. The most optimistic version of this model (for both linear and
quadratic proton acceleration) can be rejected with more than 90\% CL using
the time integrated data from 40 string configuration constraining this way
the value of the charge depletion fraction.

\section{Conclusions}
\label{sec:conc}
Searches for neutrinos in coincidence with the Sep. 2010 Crab flare have been
presented in this paper. The data used was taken with the 79-string
configuration of IceCube. This is the first analysis of data taken by this
configuration and represents the first rapid response analysis of IceCube to
an astronomical event such as the flaring of an otherwise steady standard
candle source. 
Two different approaches of event selection have been followed. One using
direct cuts on quality reconstruction variables and optimized for discovery
for $E^{-2}$ neutrino spectra, and the other based on multivariate
analysis and optimized for discovery at lower energies, important for
galactic sources that have soft spectra with cut-offs at TeV energies. The
two data sets however showed a background under-fluctuation during the time
interval considered. The corresponding upper limits based on generic neutrino
spectra have been shown for the flaring state of the Crab. 

Assuming isotropic emission from the shock (even if this may not be the case for a
highly relativistic pulsar wind) our limit for $E^{-2}$ corresponds to a
neutrino luminosity constraint for the flare state of about $\sim 2 \times 10^{35}$ erg s$^{-1}$, and $\sim 1.5 \times 10^{36}$ erg s$^{-1}$ if a neutrino cut-off of 1 TeV is assumed. In both cases the resulting neutrino luminosity constraint is about 2 -- 3 orders of magnitude lower than the spin-down luminosity of the pulsar and comparable to the peak isotropic $\gamma$-ray luminosity $\sim 5 \times 10^{35}$ erg s$^{-1}$ measured by AGILE~\cite{agile_science} in the energy range from 0.1 to 10 GeV.

In addition to the flare analysis we calculated the current best limits set
by IceCube on different models for neutrino emission from the Crab
Nebula. These limits are based on the time-integrated analysis of IceCube
with the 40-string configuration of the detector. The upper regions of
the most optimistic models can be rejected with more than 90\% CL providing
useful constraints on adjustable parameters of these models. Taking the neutrino spectrum derived from the $\gamma$-ray observations from the Crab, the constraint in neutrino luminosity for the steady emission of the Crab is $\lesssim 1 \times 10^{35}$ erg s$^{-1}$ which is a factor $\sim 1.7$ larger than the luminosity in $\gamma$-rays assuming the $\gamma$-ray spectrum measured in~Ref.~\cite{HESS2006} integrated over the energy range between 400~GeV -- 40~TeV.


In the future the IceCube detector will combine datasets from different
detector configurations. When the different livetimes of the 40-string
configuration data and the full detector will be summed, the sensitivity will
improve by about a factor of five making this search more predictive.

\section{Acknowledgements}
\label{sec:ack}
We acknowledge the support from the following agencies: U.S. National Science Foundation-Office of Polar Programs, U.S. National Science Foundation-Physics Division, University of Wisconsin Alumni Research Foundation, the Grid Laboratory Of Wisconsin (GLOW) grid infrastructure at the University of Wisconsin - Madison, the Open Science Grid (OSG) grid infrastructure; U.S. Department of Energy, and National Energy Research Scientific Computing Center, the Louisiana Optical Network Initiative (LONI) grid computing resources; National Science and Engineering Research Council of Canada; Swedish Research Council, Swedish Polar Research Secretariat, Swedish National Infrastructure for Computing (SNIC), and Knut and Alice Wallenberg Foundation, Sweden; German Ministry for Education and Research (BMBF), Deutsche Forschungsgemeinschaft (DFG), Research Department of Plasmas with Complex Interactions (Bochum), Germany; Fund for Scientific Research (FNRS-FWO), FWO Odysseus programme, Flanders Institute to encourage scientific and technological research in industry (IWT), Belgian Federal Science Policy Office (Belspo); University of Oxford, United Kingdom; Marsden Fund, New Zealand; Japan Society for Promotion of Science (JSPS); the Swiss National Science Foundation (SNSF), Switzerland; A. Gro§ acknowledges support by the EU Marie Curie OIF Program; J. P. Rodrigues acknowledges support by the Capes Foundation, Ministry of Education of Brazil.





\bibliographystyle{model1a-num-names}
\bibliography{<your-bib-database>}

\end{document}